\documentclass[12pt]{article}
\input epsf
\usepackage[dvips]{graphicx}
\usepackage{epsfig}
\begin{document}

\centerline{\Large {\bf Virtual Compton Scattering}}
\vskip 1 cm
\centerline{H. FONVIEILLE }
\vskip 5 mm
\centerline{for the Jefferson Lab Hall A and VCS Collaborations}
\vskip 5 mm
\centerline{\it Laboratoire de Physique Corpusculaire IN2P3-CNRS}
\centerline{\it Universit\'e Blaise Pascal Clermont-II, F63177 Aubi\`ere Cedex, France }

\begin{abstract}{Virtual Compton Scattering off the proton:
\ $\gamma^* p \to \gamma p$ \ is a new field of
 investigation of nucleon structure. 
Several dedicated experiments have been 
performed at low c.m. energy and various
momentum transfers, yielding specific
information on the proton. This talk reviews
the concept of nucleon Generalized Polarizabilities 
and the present experimental status.}
\end{abstract}

Virtual Compton Scattering (VCS) off the proton:
\ $\gamma^* p \to \gamma p$ \ has emerged 
within the last ten years as a powerful tool 
to study the internal structure of the nucleon, 
bringing forward exciting new concepts 
and observables. The field can be subdivided
according to different c.m. energy domains.

At high energy ($s \gg M_N^2$) and 
large momentum transfer ($ Q^2 \gg M_N^2$), 
QCD factorization allows the use of the VCS process to access:
i) the nucleon Generalized Parton Distributions (GPDs) 
via Deep VCS at small $t$~\cite{diehl} ; 
ii) the nucleon Distribution Amplitudes via 
hard Compton scattering at large $t$~\cite{da}.

At low energy, the VCS process gives access to the nucleon 
Generalized Polarizabilities or GPs, which are
the main focus of this talk. The basic concepts are introduced
and an experimental review is given of the experiments performed
in the threshold regime ($\sqrt{s} \le (M_N + M_{\pi}$)) 
and resonance region.

\section{Concept of Generalized Polarizabilities}

These are the generalization to a non-zero $Q^2$ 
of the polarizabilities introduced in 
Real Compton Scattering (RCS). 

Let us recall that polarizabilities measure
how much the internal structure of a composite particle
is deformed when an external EM field is applied.
The description of the RCS process at low energy 
involves six polarizabilities 
($\alpha,\beta$, and $\gamma_i, i=1,2,3,4$) 
which parametrize the unknown part of nucleon structure,
due to its internal EM deformation 
(see ref.3 and references therein).
Performed since the sixties, RCS experiments have
measured the proton electric and magnetic 
polarizabilities $\alpha_E$ and $\beta_M$~\cite{olmos},
the results of which can be summarized as:
i) the proton is 
a rather rigid object (polarizabilities are small), 
ii) $\beta_M$ is smaller than $\alpha_E$, due to a cancellation
between its paramagnetic and diamagnetic contributions.

These observables can be generalized to any $Q^2$, such that 
in VCS at low energy one probes the nucleon polarizability
locally inside the particle, with a distance scale given by $Q^2$.
Equivalently, VCS can be seen as elastic scattering on
a nucleon placed in an applied EM field, and hence 
the GPs can be seen as a measurement of ``distorted''
form factors. In all cases these observables are intrinsic
characteristics of the particle, and provide 
a valuable and original test of models describing 
nucleon structure.

\subsection{Photon electroproduction amplitude}

\begin{figure}[t]
\epsfxsize=11.7cm
\epsfysize=3.cm
\centerline{\epsfbox{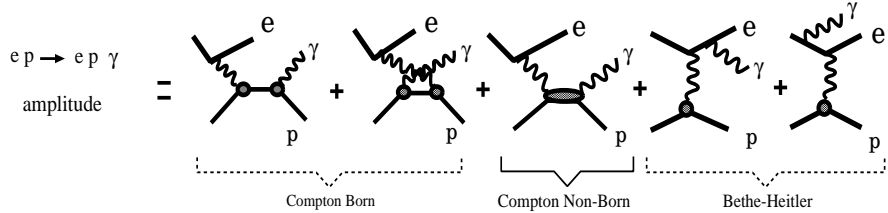}}   
\caption{Decomposition of the photon electroproduction amplitude 
into: Born ($s$ and $u$ channels), Non-Born, and 
BH contributions. \label{fig01}}
\end{figure}

\begin{figure}[t]
\epsfxsize = 11.cm
\epsfysize = 2.5cm
\centerline{\epsfbox{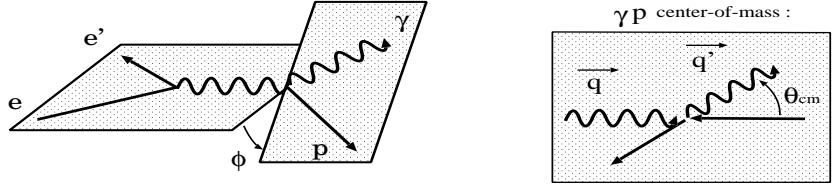}}
\caption{VCS kinematic variables. \label{fig02}}
\end{figure}

VCS is accessed by photon electroproduction 
\ $ep \to ep \gamma$ , which is the coherent sum of
the Compton process and the Bethe-Heitler (BH) process,
or electron bremsstrahlung; see Fig. \ref{fig01}.
The main kinematic variables are defined in 
Fig. \ref{fig02}:
the initial and final photon three-momenta $q$ and
$q'$ and the final
photon angles $\theta_{cm}$ and $\phi$,
in the $(\gamma p)$ center of mass.
The \ $(ep \to ep \gamma)$\ kinematics is fully
defined by these four variables plus the virtual
photon polarisation $\epsilon$.

The low-energy behavior of the amplitude $T^{ee \gamma}$
has first been worked out by P. Guichon {\it et al}~\cite{guich95}.
Only a brief summary is given here, and more details 
can be found in ref.5.
The Compton amplitude decomposes into a Born term, 
characterized by a proton in the intermediate state,
and a Non-Born term containing all other intermediate states. 
The BH and Born amplitudes are entirely
calculable, with proton EM form factors as inputs.
The Non-Born amplitude $T_{NB}$ contains the unknown part 
of the nucleon structure.
Below pion threshold a Low Energy Theorem (LET) allows to
expand $T_{NB}$ in powers of $q'$ ; 
the first term of the expansion
is a known analytical function of six independent GPs, which are
the goal of the measurements.

These GPs are derived from a multipole expansion of $T_{NB}$.
One defines mutipole amplitudes \ $H_{NB}^{(\rho ' l' \rho l )S}$ \ 
according to the two involved EM transitions:
$\rho$ ($\rho '$) stands for initial (final) 
photon polarization states
($0,1,2=$ longitudinal, magnetic, electric), $l$ ($l'$) is the
total angular momentum of the initial (final) transition, and 
$S=(0),1$ stands for proton (non)spin-flip.
The $H_{NB}$ multipoles depend on photon momenta $q$ and
$q'$ .
The GPs are defined (up to dimensional factors)  as 
the limit of $H_{NB}$ when $q' \to 0$ , i.e. in the limit
of a static EM field; they are denoted $P^{(\rho ' l' \rho l )S}$ \
and depend only on $q$.
Table \ref{tab1} summarizes the notations for the two scalar ($S=0$)
and the four spin GPs ($S=1$),  and also shows their continuity 
to the real photon point ($Q^2=0$).

\begin{table}[t]
\caption{The six lowest order Generalized Polarizabilities 
(also called dipole GPs, due to $l'=1$).\label{tab1} }
\begin{center}
\footnotesize
\begin{tabular}{|c|c|c|c|c|c|}
\hline
final $\gamma$ & initial $\gamma^*$ & S  {\small proton} &
$P^{(\rho ' l ' \rho l )S}${\small $(q)$}
& $P^{X \to Y}$ & $Q^2=0$ \\
\ & \ & {\small spin-flip} & \ & \ & {\small RCS limit} \\
\hline
E1 & C1 & 0 &  $P^{(01,01)0}$ & $P^{ C1 \to E1}$ &
${-4 \pi \over e^2} \sqrt{ {2 \over 3} }$  $ \alpha$ \\
E1 & C1 & 1 &  $P^{(01,01)1}$ & $P^{ C1 \to E1}$ &
0 \ \ \ \ \\
M1 & M1 & 0 &  $P^{(11,11)0}$ & $P^{ M1 \to M1}$ &
${-4 \pi \over e^2} \sqrt{ {8 \over 3} }$  $ \beta$ \\
M1 & M1 & 1 &  $P^{(11,11)1}$ & $P^{ M1 \to M1}$ &
0 \ \ \ \ \\
E1 & M2 & 1 &  $P^{(01,12)1}$ & $P^{ M2 \to E1}$ &
${-4 \pi \over e^2} {\sqrt{2} \over 3} \gamma_3$ \ \ \ \ \\
M1 & C2 & 1 &  $P^{(11,02)1}$ & $P^{ C2 \to M1}$ &
${-4 \pi \over e^2} \sqrt{ {8 \over 27}}
(\gamma_2 + \gamma_4)$ \\
\hline
\end{tabular}
\end{center}
\end{table}

Below pion threshold, the VCS amplitude is purely real;
above pion threshold it becomes complex, and resonances can be
produced on-shell. One may say that the GPs 
are conceptually linked to the
contribution of virtual resonant intermediate states,
extrapolated down to $q'=0$ .

\subsection{Photon electroproduction cross section}

The photon electroproduction cross section is evaluated
from the relation: \
 $ \vert  T^{e e \gamma} \vert ^2 \ = \ 
   \vert T_{BH+Born} \vert ^2  \ + \ 
2 \mbox{Re} (T_{BH+Born} \times T_{NB}) \ + \
\vert T_{NB} \vert ^2$ . Below pion threshold, 
\ $\vert T_{NB} \vert ^2$ \ can be neglected and thus
the GPs are extracted via the interference term (BH+Born)(NB).
The LET leads to the following expression for the unpolarized
cross section:
\footnote{ $d \sigma$ is  a short notation for the
fivefold differential cross section \ $d^5 \sigma / 
d \Omega_{e'}^{lab} \ dk'^{lab} \ d \Omega_{p'}^{cm}$.}
\begin{eqnarray} \begin{array}{l}
d \sigma (ep \gamma) \  = \  d \sigma_{BH+Born} \  + \\ 
\ \ \ \ \ (P.S.) \times {\bigg [} \ 
v_1 {\big (} 
\ P_{LL}(q)- {1 \over \epsilon} P_{TT}(q) \ {\big )} \ + \
v_2 {\big (} \ P_{LT}(q) \ {\big )} \ 
{\bigg ]} \ + \ O(q'^2) \ , \\
\end{array} \label{eq01} \end{eqnarray}
where $(P.S.)$ is a phase space factor and 
$v_1,v_2$ are known kinematic coefficients~\cite{pgvdh98}.
The two structure functions 
$ (P_{LL}- {1 \over \epsilon} P_{TT})$ \ and \ $ P_{LT}$ \ 
are linear combinations of the GPs, given e.g. by the 
following choice:~\cite{pgvdh98}

\begin{eqnarray*} \begin{array}{lll}
 P_{LL}(q) & = & -2 \sqrt{6} \ M_N \ G_E(\tilde Q ^2 ) \
 P^{(01,01)0}(q) \\
 P_{TT} (q) & = & -3 \ G_M(\tilde Q ^2 ) \
{ q^2 \over {\tilde q _0} }  \times
{\bigg [ } \
 P^{(11,11)1}(q)  - \sqrt{2} \tilde q _0
\ P^{(01,12)1}(q)
\ {\bigg ] }  \\
 P_{LT}(q) & = & \sqrt{ {3 \over 2} } \
{ M_N \ q \over {\tilde Q} } \ G_E(\tilde Q ^2 ) \
P^{(11,11)0}(q) +
{3 \over 2} \ { {\tilde Q} \ q \over {\tilde q _0} } \
G_M(\tilde Q ^2 ) \
P^{(01,01)1}(q) \ , \\
\end{array}  \end{eqnarray*}
where $ {\tilde Q ^2}, \ {\tilde q _0}, \ {\tilde Q}$ 
are specific kinematic variables
\footnote{ The important notion is that $ {\tilde Q ^2}$
is equivalent to $q$.}.
So  in an unpolarized experiment performed 
at fixed $q$ and $\epsilon$, one measures 
two structure functions:  
$ (P_{LL}- {1 \over \epsilon} P_{TT})$ sensitive to
the electric GP \ $\alpha(Q^2) \sim P^{(01,01)0} $ ,
 and \ $ P_{LT}$ \ sensitive to the magnetic GP \ 
$\beta(Q^2) \sim P^{(11,11)0} $ .

\subsection{Methods to extract GPs} \label{sec01}

Two methods are presently used to extract GPs from
absolute $(ep \gamma)$ cross sections.  \newline
$\bullet$ {\bf Method 1}  is based on the LET,
and only works below pion threshold.
In bins of photon angles 
$(\theta_{cm}, \phi)$,
one forms the quantity \
$( d \sigma_{exp} - d \sigma_{BH+Born})/(P.S.)$ \
measured at finite $q'$, and extrapolates it to
$q'=0$ to obtain the term in brackets in eq. \ref{eq01}. 
Present experimental data suggest that, 
at least in most of the phase space,
the extrapolation can be done assuming that
the $O(q'^2)$ contribution in eq. \ref{eq01} is negligible.
The bracketed term is then easily fitted as a linear combination of 
the two structure functions \ 
$ (P_{LL}- {1 \over \epsilon} P_{TT})$ \  and \ $ P_{LT}$ \
at fixed $q$ and $\epsilon$. \newline
$\bullet$ {\bf Method 2} is based on the formalism of 
Dispersion Relations (DR)~\cite{barbara} and works below
pion threshold as well as in the first resonance region. 
In this model the imaginary part 
of the VCS amplitude is given by the sum of 
$\pi N$ intermediate states, 
computed from \ $\gamma^* N \to \pi N$ \  data (MAID model),
plus higher order contributions which are not constrained
by the model. The latter have to be fitted to the VCS data,
under the form of two free parameters \ $\Lambda_{\alpha}$ \
and \ $\Lambda_{\beta}$ \ describing the $Q^2$-dependence
of the scalar GPs \ $\alpha$ and $\beta$. The knowledge of the
parameters at a given value of $Q^2$ then yields the model 
prediction for the structure functions 
$ P_{LL}$ , $P_{TT}$ \  and \ $ P_{LT}$ 
at this momentum transfer.

\subsection{GP effect on cross sections}

\begin{figure}[t]
\epsfxsize = 12.cm
\epsfysize = 9.cm
\centerline{\epsfbox{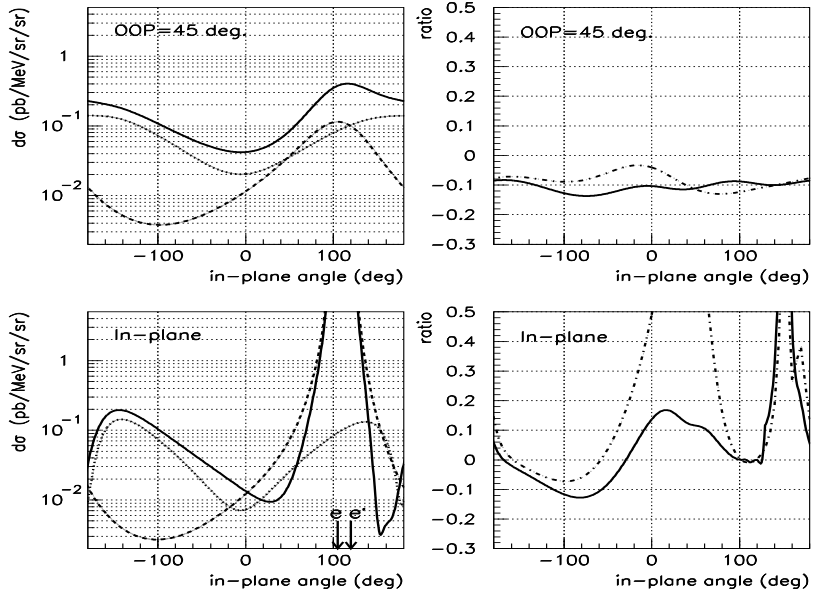}}
\caption{ $(ep \to ep \gamma)$ cross section components
for: $q=1.08$ GeV/c, $ q'=$ 105 MeV/c, and 
$\epsilon=0.95$ .
The abscissa is the azimuthal angle of the outgoing
photon when the  polar axis is chosen perpendicular to the 
leptonic plane. ``OOP'' is the polar angle using 
this convention. Left plots: BH+Born (solid), BH (dashed),
and Born (dotted) contributions. Right plots: the ratio \  
$( d \sigma_{GP} - d \sigma_{BH+Born})/ d \sigma_{BH+Born}$ \  
for two calculations of $  d \sigma_{GP}$ : 
i) a first order GP effect, taking 
$P_{LL}-{1 \over \epsilon}P_{TT}=2.3$ GeV$^{-2}$ and
$P_{LT}=-0.5$ GeV$^{-2}$ (solid), ii)
a DR calculation for parameter values
$\Lambda_{\alpha}=$ 0.92 GeV,
$\Lambda_{\beta}=$ 0.66 GeV (dash-dotted).\label{fig03}}
\end{figure}

Figure \ref{fig03}-left shows the various components of the photon
electroproduction cross section, in and out of the leptonic
plane, for selected kinematics. The Bethe-Heitler peak is dominant
around the incident and scattered electron directions; 
as one goes out-of-plane it fades away, giving a 
smoother cross section behavior.
Figure \ref{fig03}-right shows the expected 
effect of GPs on the cross section,
as given by two different calculations:
the lowest order (or bracketed) term of eq. \ref{eq01}, and the full 
DR prediction. Out-of-plane, the GP effect is roughly constant,
of the order of -10 \%. In-plane the GP effect has a more
complicated pattern, due to the BH interference.

\section{Experiments}

Table \ref{tab2} summarizes the VCS experiments performed so far.
All of them have detected the scattered electron and 
outgoing proton in high-resolution magnetic spectrometers,
selecting the exclusive photon channel by the missing-mass technique.
Also, being unpolarized experiments, they all measure the same
two structure functions,
at different values of $q$. An accurate determination 
of the absolute five-fold cross sections is necessary, 
due to the relatively small polarizability effect.

\begin{table}[t]
\caption{VCS experiments. \label{tab2}}
\begin{center}
\footnotesize
\begin{tabular}{|c|c|c|c|c|c|}
\hline
experiment & $Q^2$ & $(\gamma^* p)$ c.m.
& $p$ cone & data &  status \\
$e p \to e p \gamma$ &  (GeV$^2$) & energy $\sqrt{s}$ & 
$\theta_{pq \ lab}$ & taking & {\tiny (@ end 2001)} \\
\hline
 MAMI \ A1 & 0.33   & $< (M_N + M_{\pi})$ &
10$^{\circ}$ &  1995+97 & published \\
 JLab E93-050 &  1.0, 1.9 & $< 1.9 $ GeV &
 6$^{\circ}$, 3$^{\circ}$ & 1998 &  final stage \\
 Bates \ E97-03 & 0.05    & $< (M_N + M_{\pi})$ &
 28$^{\circ}$  OOPS & 2000 & analysis \\
 Bates \ E97-05 & 0.12  & $ \sim 1.232$ GeV    &
 14-20$^{\circ}$  OOPS
& 2001 &  analysis \\
\hline
\end{tabular}
\end{center}
\end{table}

\subsection{The MAMI experiment}

\begin{figure}[t]
\begin{center}
\epsfxsize = 10.cm
\epsfysize = 4.5cm
\centerline{\epsfbox{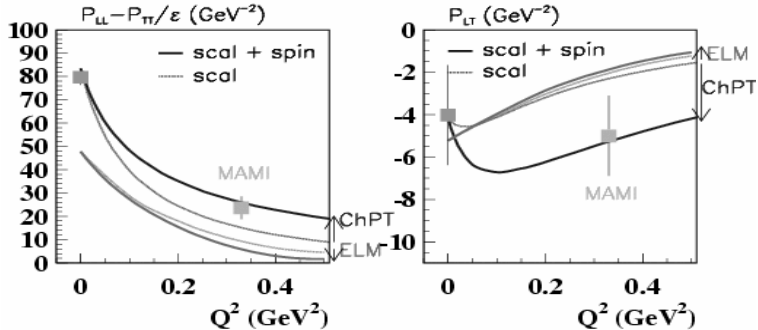}}
\end{center}
\caption{VCS unpolarized structure functions measured
at Mainz~\cite{jroche}
and their value at the real photon point~\cite{olmos}. 
The curves represent two model predictions (CHPT~\cite{hemmert}
and ELM~\cite{vdhelm}), including (in dark grey) or not
including (in light gray) the spin GPs (effect indicated by an arrow).
\label{fig04}}
\end{figure}

The Mainz experiment~\cite{jroche} measured
photon electroproduction cross sections 
in the leptonic plane, at $Q^2=0.33$ GeV$^2$. The two
structure functions \ $P_{LL}-P_{TT}/\epsilon$ \ and \ $P_{LT}$ \
were determined using the LET method as described
in section \ref{sec01}, at $q=0.6$ GeV/c and $\epsilon=0.62$.
Results are plotted in Fig. \ref{fig04}; 
they show good agreement with the calculation of 
Heavy Baryon Chiral Perturbation Theory~\cite{hemmert}.
Several models predict an extremum of \ $P_{LT}$ \ at low $Q^2$,
a feature which will be interesting to confirm experimentally. 
This turnover can be related to the behavior of the 
para- and diamagnetic contributions to the
$\beta$ polarizability. In CHPT it originates from 
the pion cloud, which yields a diamagnetic contribution of
positive sign, visible at low  $Q^2$.
For a review of  model predictions see ref.9.

\subsection{The BATES experiments}

The Bates experiment 97-03~\cite{bates01}
has been performed at $Q^2=0.05$ GeV$^2$, i.e. in the
region of the expected turnover of $P_{LT}$.
Measurements have been done in-plane and at 90$^{\circ}$
out-of-plane, using the OOPS spectrometers. The experiment 
covers a limited range in polar angle 
$\theta_{cm}$ around 90$^{\circ}$, so
the structure functions will be extracted mostly from the
$\phi$-dependence of the cross section.
Data analysis is in progress,  presently concentrating 
on Monte-carlo studies and absolute normalization.
This experiment
represents a Lab achievement, having made the first use of
the high duty factor beam in the South Hall Ring and of the 
full OOPS system.
The Bates experiment 97-05~\cite{bates02} has been performed at 
$Q^2=0.12$ GeV$^2$ to study the $N \to \Delta$ transition, and
data analysis is also in progress.

\subsection{The JLab experiment}

Experiment E93-050~\cite{jlabvcs} was performed in Hall A of the
Thomas Jefferson National Accelerator Facility (JLab)
at $Q^2=$ 1.0 and 1.9 GeV$^2$.
Data covers the region below pion threshold, and the resonance
region up to $\sqrt{s}=$ 2 GeV at $Q^2=$ 1.0 GeV$^2$.

The strong Lorentz boost from $\gamma p$ center-of-mass to lab
focuses the outgoing proton in a narrow cone (see Table \ref{tab2})
allowing the hadron arm acceptance to cover the full phase space 
of the outgoing photon in c.m.
The key points to obtain accurate cross sections
are a detailed Monte-carlo 
simulation (including radiative corrections) and
a detailed study of cuts in order to eliminate background, 
mainly due to punchthrough protons.

Absolute normalization is checked in two ways:
i) by computing the $(ep \to ep)$ cross section from 
elastic data taken during the experiment;
ii) using the VCS data, namely the important property
that the $(ep \gamma)$ cross section should tend to 
the known (BH + Born) cross section
when the final photon momentum $q'$ tends to zero
\footnote{indeed, in eq. \ref{eq01} the bracketed term is of order
$(q')^0$ and the phase space factor $(P.S.)$ is of order $(q')^1$
.}. Both tests show that the absolute normalization
is correct within 1-2 percent, when using
the most recent determination of proton form factors:
the JLab measurement of the ratio $\mu G_E/G_M$~\cite{gayou}
and the $G_M$ fit of ref.15.

\par
\begin{figure}[t]
\epsfxsize = 12.cm
\epsfysize = 10.cm
\centerline{\epsfbox{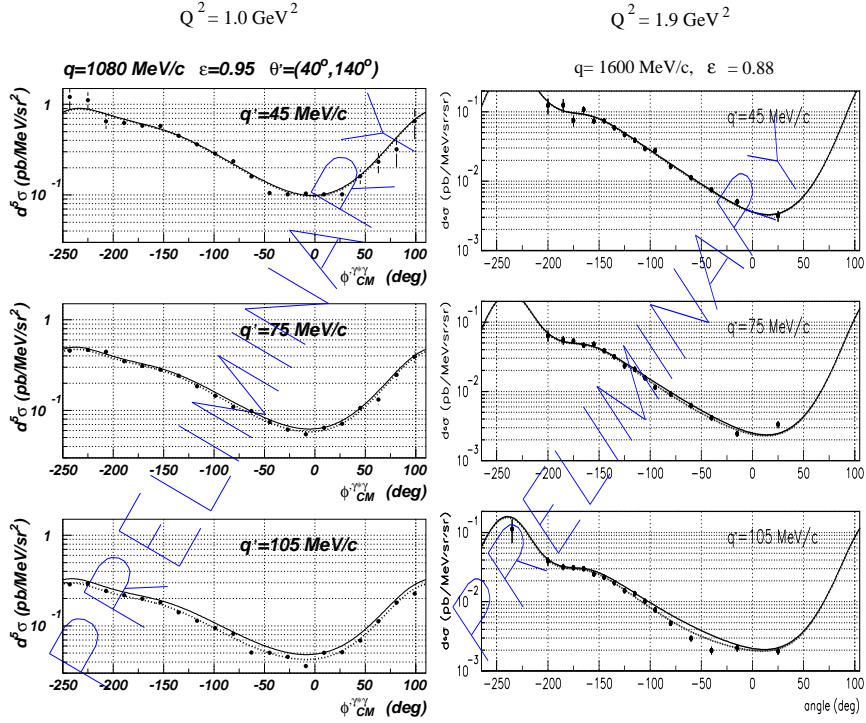}}
\caption{$(ep \to ep \gamma)$  cross sections measured at JLab  
versus in-plane angle, for OOP= 50$^{\circ}$ (left)
and 25$^{\circ}$ (right). Error bars are statistical only. 
The angle in abscissa is the same as in Fig. 3. 
The curves correspond to: BH+Born calculation (solid)
and a first order GP effect (dotted). 
\label{fig05}}
\end{figure}

\noindent
$\bullet$ {\bf Analysis below pion threshold:}
photon electroproduction cross sections 
have been obtained at fixed $q=1.08 (1.60)$  GeV/c and 
fixed $\epsilon=0.95 (0.88)$, corresponding to
the data set at $Q^2=$ 1.0 (1.9) GeV$^2$.
As an example, Fig. \ref{fig05} shows some of the 
out-of-plane cross sections measured for both data sets.
These data illustrate how the (small) GP effect 
increases with $q'$ and how its shape agrees
with the LET prediction. More details can be found 
in ref.17.

The quantity \ $( d \sigma_{exp} - d \sigma_{BH+Born})/(P.S.)$ \
of eq. \ref{eq01} does not show any noticeable $q'$-dependence,
so it is averaged over  $q'$ and then fitted according
to the first method of section \ref{sec01}.
The fit is performed on (in-plane + out-of-plane) cross sections 
and gives a reasonably good $\chi^2$,
confirming the validity of the low-energy expansion at these 
rather high $Q^2$. Numerical results are reported in Table 
\ref{tab3}.
A second analysis of this data below pion threshold
is presently underway, based on the DR model. A preliminary 
result at $Q^2=1.9$ GeV$^2$ is included in Table \ref{tab3}.

\par
\begin{figure}[t]
\epsfxsize = 12.cm
\epsfysize = 12.cm
\centerline{\epsfbox{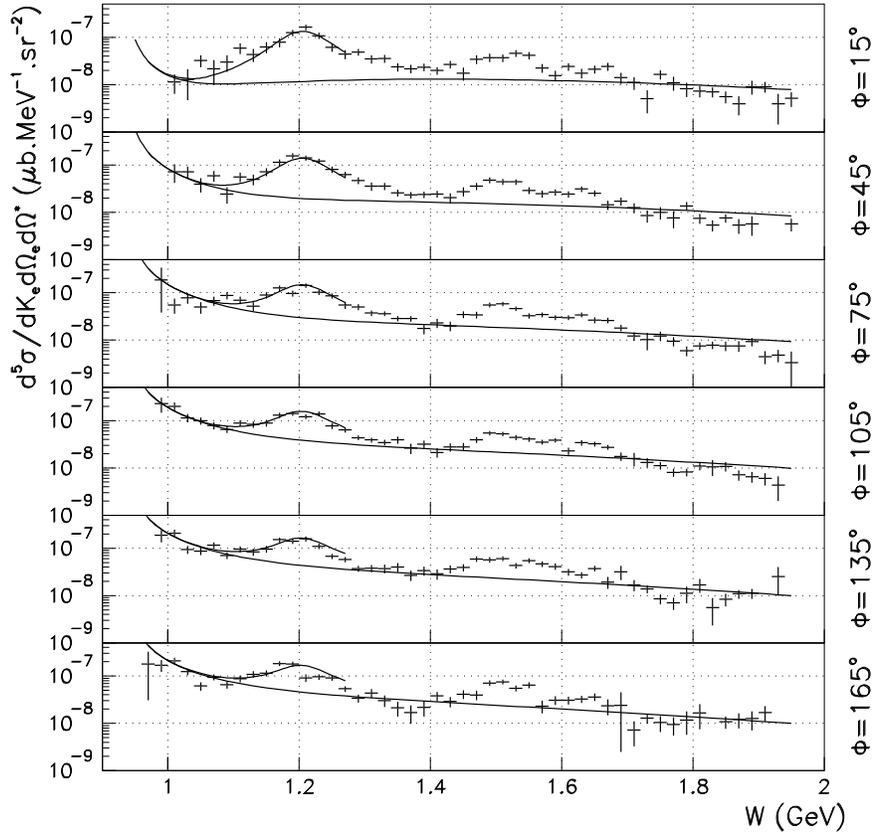}}
\caption{Photon electroproduction cross sections
in the resonance region. The curve spanning 
the whole range in $W$ is the BH+Born calculation. The curve
limited to $W<1.25$ GeV is the DR prediction for parameter values
$\Lambda_{\alpha}= 1.0$ and $\Lambda_{\beta}= 0.45$ GeV.
\label{fig06}}
\end{figure}

\noindent
$\bullet$ {\bf Analysis in the resonance region:}
these are the first VCS measurements ever performed in
this kinematic domain. The initial goal was to study how 
resonances couple to the doubly EM channel, and search 
for possible missing resonances. Doing an excitation scan 
in $W=\sqrt{s}$ from $M_N$ to 1.9 GeV, 
cross sections have been determined 
at a fixed $Q^2=1.0$ GeV$^2$, 
backward angle $\theta_{cm}=167.2^{\circ}$ and beam energy
4.032 GeV~\cite{resonan}. 
They are presented in Fig. \ref{fig06} as a function of $W$
for various azimuthal angles $\phi$.
The DR model reproduces well the Delta region.
Using these data, the second method of
section \ref{sec01} has been applied for the first time.
The free parameters of the DR model are adjusted 
by a $\chi^2$ minimization, yielding  
the values of the two structure functions
at $Q^2=1.0$ GeV$^2$ and $\epsilon=$0.95.
Results are reported in Table \ref{tab3}.

\subsection{Results summary}

Table \ref{tab3} summarizes our present knowledge of the
two structure functions measured in unpolarized VCS:
the MAMI result, and the preliminary JLab results 
obtained so far, both below and above pion threshold.
One first notices the fast decrease of the observables
with $Q^2$, similarly to form factors. Second, for the 
JLab data there is a nice agreement between the results
obtained by the two methods, LET and DR. 
These new measurements should stimulate theoretical 
calculations of GPs at high $Q^2$. Indeed most model 
predictions are presently limited to $Q^2 \ll 1$ GeV$^2$.

\begin{table}[t]
\caption{Present results on VCS structure functions. 
$(\pm \bullet )$ indicate that systematic errors are not fully determined.
DR results at $Q^2=1.0$ GeV$^2$ are from  
resonance region analysis.
\label{tab3}}
\begin{center}
\footnotesize
\begin{tabular}{|l|l|l|lll|}
\hline
$Q^2$ & $q$ & $\epsilon$ & 
\multicolumn{3}{|c|}{  $P_{LL}-P_{TT}$ structure function
 {\small (GeV$^{-2}$)} } \\
 {\tiny (GeV$^2$)} &{\tiny (MeV/c)} &   & & &  \\ 
\hline
0.33 & 600 & 0.62 &
 LET: \ + 23.7 & $\pm$ 2.2 {\tiny (stat)} & 
$\pm$ 4.3 {\tiny (syst)}  \\
\hline
 1.0  &  1080  & 0.95 &
 LET: \ +2.32 & $\pm$ 0.22  {\tiny (stat)}
 & $\pm$ 0.35 {\tiny  (syst)} \ $\pm \bullet$ \\
1.0 & 1133 & 0.95 & 
 DR : \ + 2.29 & $\pm$ 0.24 {\tiny  (stat)} & 
$^{-0.49}_{+0.30}$ \ {\tiny  (syst)}  \\
\hline
1.9  &   1600  & 0.88 &
 LET: \ + 0.56 & $\pm$ 0.07 {\tiny  (stat)} 
& $\pm$ 0.11 {\tiny  (syst)} \ $\pm \bullet$  \\
1.9 & 1600 & 0.88 & 
 DR :  in [+0.43,+0.84] & &  \\
\hline \hline
$Q^2$ & $q$ & $\epsilon$ & 
\multicolumn{3}{|c|}{  $P_{LT}$ structure function
 {\small (GeV$^{-2}$)} } \\
 {\tiny (GeV$^2$)} &{\tiny (MeV/c)} &   & & &  \\ 
\hline
0.33 & 600 & 0.62 &
 LET: \ $-$ 5.0 & $\pm$ 0.8 {\tiny (stat)} & 
$\pm$ 1.8 {\tiny (syst)}  \\
\hline
 1.0  &  1080  & 0.95 &
 LET: \ $-$ 0.42 & $\pm$ 0.11  {\tiny (stat)}
 & $\pm$ 0.02 {\tiny  (syst)} \ $\pm \bullet$ \\
1.0 & 1133 & 0.95 & 
 DR : \ $-$ 0.53 & $\pm$ 0.12 {\tiny  (stat)} & 
$^{-0.03}_{+0.16}$ \ {\tiny  (syst)}  \\
\hline
1.9  &   1600  & 0.88 &
 LET: \ +0.009 & $\pm$ 0.041 {\tiny  (stat)} 
& $\pm$ 0.005 {\tiny  (syst)} \ $\pm \bullet$  \\
1.9 & 1600 & 0.88 & 
 DR :  in [$-$0.05,+0.02] & & \\
\hline
\end{tabular} 
\end{center}
\end{table}

\section{Future prospects}

Polarizabilities and Generalized Polarizabilities
are intrinsic characteristics of composite particles. As such 
it is interesting to measure them for every hadron, including e.g.
the neutron, pion, etc, although this may seem a task 
for the far future.
Undoubtedly, investigations of the VCS process on the proton
have been fruitful, and will continue to bring exciting new
results in the coming years. Future developments are foreseen 
in various c.m. energy domains: \newline
$\bullet$ at low energy, the mapping of the unpolarized
structure functions  $(P_{LL}-{1 \over \epsilon} P_{TT})$
and $P_{LT}$ versus $Q^2$ can be completed,
and $P_{LL}$ and $P_{TT}$ can be disentangled by an
 $\epsilon$-separation. By studying \ 
$\vec e p \to e \vec p \gamma $ \
with a  polarized beam and recoil proton polarimetry, 
one can in
principle disentangle the six independent GPs entering the
first order term of the LET, giving access to the spin GPs.
Such experiments are planned at Mainz and Bates. \newline
$\bullet$ In the resonance region, the VCS process was
investigated for the first time by the JLab E93-050 experiment.
It demonstrated the feasibility of GP extraction
above pion threshold,
owing to an enhanced sensitivity of the VCS cross section 
to GPs in the Delta resonance region.
Future experiments along these lines at high $Q^2$ are 
foreseeable, either with or without polarization
degrees of freedom. \newline
$\bullet$ VCS at higher energies is certainly a very 
active field, with growing interest in Deep VCS and
the GPDs.
Using a longitudinally polarized electron beam, 
the HERMES and JLab-CLAS collaborations have
determined a Single Spin Asymmetry
in \ $\vec e p \to e p \gamma$~\cite{dvcs}, giving the first input to
GPD models, and more such experiments are planned
at JLab, HERMES and COMPASS in the near future.
As energies increase, experimental resolution limitations make
it more and more difficult to isolate the 
one-photon electroproduction channel, and all present
and future experiments  plan to detect all three particles 
in the final state in order to reach exclusivity.

\section*{Acknowledgments }

This work was supported by DOE, NSF, by contract DE-AC05-84ER40150
under which the Southeastern Universities Research Association
(SURA) operates the Thomas Jefferson National Accelerator Facility
for DOE, by the French CEA, the UBP-Clermont-Fd and CNRS-IN2P3 (France),
the FWO-Flanders and the BOF-Gent University (Belgium) and by the
European Commission ERB FMRX-CT96-0008.



\begin{thebibliography}{999}
\bibitem{diehl}{ M. Diehl, these Proceedings.}
\bibitem{da}{ M. Vanderhaeghen {\it et al.}, Nucl. Phys. A622 (1997) 144c.}
\bibitem{olmos}{ V. Olmos de Leon {\it et al.}, Eur. Phys. J. A10 (2001) 207.}
\bibitem{guich95}{ P. Guichon {\it et al.}, Nucl.Phys. A591 (1995) 606.}
\bibitem{pgvdh98}{ P. Guichon {\it et al.}, Prog. Part. Nucl. Phys. 41 (1998) 125.}
\bibitem{barbara}{ B. Pasquini {\it et al.}, Eur. Phys. J. A11 (2001) 185,
and these Proceedings.}
\bibitem{jroche}{ J. Roche {\it et al.}, Phys. Rev. Lett. 85 (2000) 708.}
\bibitem{hemmert}{ T. Hemmert {\it et al.}, Phys. Rev. D55 (1997) 2630.}
\bibitem{revmod}{ See ref. [8], [10], [6], and also:
A. Metz {\it et al.}, Z. Phys. A356 (1996); 
G. Liu {\it et al.}, Aust. J. Phys. 49 (1996);
B. Pasquini {\it et al.}, nucl-th/0105074.}
\bibitem{vdhelm}{ M. Vanderhaeghen, Phys. Lett. B368 (1996) 13.}
\bibitem{bates01}{ J. Shaw, R. Miskimen {\it et al.}, 
Bates Proposal E97-03 (1997).}
\bibitem{bates02}{ N. Kaloskamis, C. Papanicolas {\it et al.}, 
Bates Proposal E97-05 (1997).}
\bibitem{jlabvcs}{ P. Bertin {\it et al.}, JLab proposal E93-050 (1993).}
\bibitem{gayou}{ O. Gayou {\it et al.}, Phys. Rev. Lett. 88 (2002) 092301.} 
\bibitem{brash}{ E. Brash {\it et al.}, Phys. Rev. C65 (2002) 051001.}   
\bibitem{resonan}{G. Laveissi\`ere, Thesis DU 1309, UBP Clermont-Fd (2001),
and also L. Todor, these Proceedings.}
\bibitem{luc}{ L. Van Hoorebeke, these Proceedings.}
\bibitem{dvcs}{ S. Stepanyan {\it et al.}, 
Phys. Rev. Lett. 87 (2001) 182002;
A. Airapetian {\it et al.}, 
Phys. Rev. Lett. 87 (2001) 182001.} 
\end{thebibliography}
\end{document}